\documentclass[12pt]{article}
\usepackage{amsmath}
\usepackage{setspace}
\title{Creation of a Toy Universe}
\author{Eric G. Novak\thanks{novak@ihes.fr}\\
\emph{Institut des Hautes \'{E}tudes Scientifiques, 35 Route de Chartres}\\
\emph{91440 Bures-sur-Yvette, France}}

\begin{document}
\maketitle
\begin{abstract}
General ideas of gauge/gravity duality allow for the possibility of time dependent solutions that 
interpolate between a perturbative gauge theory phase and a weakly curved string/gravity phase.
Such a scenario applied to cosmology would exhibit a non-geometric phase before the big bang.
We investigate a toy model for such a cosmology, whose endpoint is the classical limit of the
two-dimensional non-critical string.
We discuss the basic dynamics of this model, in particular how it evolves toward the double scaling 
limit required for stringy dynamics.  We further comment on the physics that will determine the fluctuation
spectrum of the scalar tachyon.
Finally, we discuss various features of 
this model, and what relevance they might have for
a more realistic, higher dimensional scenario.
\end{abstract}

\section{Introduction}
The principle of holography \cite{'tHooft:1993gx,Susskind:1994vu, Bousso:2002ju} seems to imply
that space-time 
geometry emerges dynamically from more fundamental degrees of freedom.
In the AdS/CFT correspondence \cite{Maldacena:1997re, Aharony:1999ti}, for example, it is the CFT that
enjoys a more fundamental, non-perturbative description, while the stringy, geometric 
description is applicable
only in the regime of large 't Hooft coupling.  For time dependent solutions, it 
is possible that
the geometric description is only applicable for part of the time.  
In particular, one may consider a cosmology that will be qualitatively very different from existing proposals,
where the universe at early times does not enjoy a geometrical description.

Using the qualitative features of the AdS/CFT correspondence as a guide, we assume that the correct description
of quantum gravity has a holographic description given by the dynamics of some theory of $N \times N$ matrices.
This theory may be either a regular large $N$ gauge theory or something more general, such as a suitable open
string field theory with $N$-dimensional Chan-Paton factors.  We further assume that in the large $N$ limit this
theory has a (closed) string expansion \cite{'tHooft:1973jz}, with an effective 't Hooft coupling $\lambda_{\rm eff}
\propto N$.  When the 't Hooft coupling is small the theory enjoys a (suitably generalized) 
perturbative gauge description, while large 't Hooft coupling corresponds to a weakly curved gravitational description,
similarly to the case in AdS/CFT where the AdS radius $R$ is proportional to $\lambda_{\rm eff}^{1/4}$.
Now if we consider an expanding universe, suitably chosen holographic screens should also be expanding.  
We take a simple interpretation of holography, wherein if a comoving region of spacetime expands, the 
growing entropy limit implies a growing number of fundamental 
degrees of freedom describing the region.  Thus we are imposing that {\em locality} continues to hold in some
generalized sense.
As the universe cools and expands, an increase in the number of degrees of freedom implies a 
growth in the number $N$ characterizing the fundamental ``gauge'' description.
Thus we assume that out of some $N_{\rm tot} \times N_{\rm tot}$ total degrees of freedom
(with $N_{\rm tot}$ perhaps infinite),
and at a particular epoch $t$, all but a certain $N_{\rm eff}(t) \times N_{\rm eff}(t)$ subset of the 
degrees of freedom are frozen out from the dynamics.
The cooling of the universe then involves the thawing out of these other degrees of freedom, which then 
allow for the holographic description of the growing universe.
Since we are now, in effect, allowing for a time dependent $N$, we expect that the dynamics of the large N expansion is controlled by an effective 't Hooft coupling 
$\lambda_{\rm eff} (t)
\propto N_{\rm eff}(t)$.  As the universe cools, from a small subset of highly excited degrees of freedom
to a much larger set in the future, the 't Hooft coupling grows, and we travel from a weak gauge coupling 
regime
to a strong coupling regime.  It is the strong 't Hooft coupling regime which should have a stringy, 
geometric description.

As a potential mechanism for cosmology, we merely point out some basic ideas, leaving a more detailed 
investigation to future work \cite{future}.
\begin{itemize}
\item  We begin with a very high temperature, weakly coupled gauge theory.  This system will have some characteristic
spectrum of fluctuations.  Further, depending on the order of the transition between the gauge and gravity phases,
there will be some characteristic spectrum.  For example, in condensed matter physics, second order transitions
produce scale-invariant fluctuations in effective field theories.  These fluctuations will then naturally translate
into some set of fluctuations in the gravity description, and in particular may seed the formation of large scale
structure.  It is interesting in this respect to note that the cosmic microwave background fluctuations are indeed
very nearly scale invariant, a feature conventionally associated with slow-roll inflation.
\item  If the radius of curvature is indeed proportional to some positive power of the time dependent, effective
't Hooft coupling, as in AdS/CFT, then a mechanism like ours, driving the system 
towards large $N$, will
move towards a flat universe.  In addition, since the evolution of the universe will naturally extend
past the big bang into the gauge phase, the horizon problem can be naturally resolved.  The scenario may
thus provide an alternative to inflation.
\item  Even in the case that there is still an inflationary epoch, our scenario may be useful for understanding
the initial conditions.  That is to say that initial conditions that may be very natural for a highly excited,
perturbative gauge theory may look rather exotic by the time that they are propagated (through the dynamics of
the phase transition) into initial conditions for the geometric phase.  
\item String cosmology is typically studied using the low energy effective field theory, with perhaps some exotic
non-perturbative features, like D-branes, added to the mix.  In this context, it has become important to understand
the huge space of (apparently) allowed string vacua, which has been termed the string ``Landscape" \cite{Susskind:2003kw, Kachru:2003sx}.
In order to make solid arguments regarding the statistics of the landscape (as in the work of Douglas \cite{Douglas:2003um}), it seems
necessary to also understand the behavior of strongly coupled models (related to the gauge phase), and the motion
between strongly and weakly coupled regions.
\end{itemize}

Most modestly, then, this scenario is important for understanding the pre-inflationary phase in Landscape 
models which might differ, for example, from those that exhibit eternal inflation. Perhaps
a bit more ambitiously, it may also serve as an alternative to inflation and as an explanation for the big 
bang and large scale structure of the universe.  We leave this possibility to future work.

In this paper we make a first step towards exploring this idea within the context of the 
simplest gauge/gravity duality, that of the $c=1$ matrix model \cite{reviews}. 
Our aim in this work is to see how we would implement our scenario in this simplified set-up, and to see 
what sort of surprises may
be in store for us.  In section two we outline the basic dynamics of our model.  
In particular, we find that the ordinary $c=1$ matrix model does not allow for a suitable freezing of
 the degrees of freedom, and we are led to introduce a mechanism where
quantum effects (and thus eigenvalue repulsion) are turned on as the system cools.
We further find that the instability of the bosonic string is crucial in allowing the system
to dynamically approach the double scaling limit.
It is not clear at this point whether these modifications have importance for higher dimensional scenarios, 
or whether
they are necessitated by the special kinematics of 2D.
In section three we study the nature of the gauge/gravity phase transition in our model,
and briefly discuss how to extract the fluctuation spectrum for the tachyon.  We also discuss the most basic 
features
of the two dimensional cosmology.
We conclude in section four by comparing the two dimensional model and it's cosmology to what we may expect 
to hold in higher dimensions, pointing the way to questions that we shall address in future work.

\section{The Basic Scenario}
We begin this section with a brief review of the key features of the $c=1$ matrix model \cite{reviews},
emphasizing it's role as a simple example of gauge/gravity duality, as clarified in recent work spearheaded 
by McGreevy and Verlinde \cite{McGreevy:2003kb}.  Since there have been very many papers on this subject
recently, our treatment will be brief.
The model is described by the $U(N)$ quantum mechanics of a single $N \times N$ matrix $M$, moving in a 
$U(N)$ invariant potential $V(M)$.  In this
paper we will be concerned with the bosonic string, which is related to the matrix model with 
a cubic potential.
Following a standard recipe, one can integrate out the ``off-diagonal" components of $M$, as well as the
gauge field, finally reducing the dynamics
to those of the matrix eigenvalues, which behave as free fermions $\lambda_i$ moving independently in the
potential $V(\lambda_i)$.  The ground state of this theory
is obtained by simply filling up the fermi sea to some fermi energy $\epsilon_F$. 

The string theory phase of this system is approached by taking the {\em double scaling limit}.  One fills
the fermi sea up to a fermi energy that approaches a critical value $\epsilon_c$,
the local maximum of the cubic potential.  Introducing the
parameter $\mu = N (\epsilon_c - \epsilon_F)$, the double scaling limit involves taking the rank of the 
matrix $ N \rightarrow \infty$ and $\epsilon_c - \epsilon_F \rightarrow 0$ with $\mu$ fixed.  In this limit
the fermion sea near the local maximum describes a two dimensional linear dilaton string background.
The scalar ``tachyon" (which is actually massless in two dimensions) is then related to the collective field
describing small fluctuations of the Fermi sea.

Some time dependent solutions of this model, proposed as solutions for two-dimensional cosmology, were 
discussed by Karczmarek and Strominger \cite{Karczmarek:2003pv}.  In these models, an incoming sea of 
fermions is chosen so that as the fermions move towards the maximum of the potential, they form a smooth
region of spacetime for some finite time period.  This picture is qualitatively very different from what 
we want to produce, as the double scaling limit is implicitly taken in the initial conditions.  Namely one
needs to choose a large enough blob of eigenvalues ($N$ large), with momentum tuned precisely so that they
get sufficiently close to the critical point of the potential.

What we want to have in our case is a system that dynamically approaches the double 
scaling limit, as the total energy spreads from an initial small subset of highly excited degrees of freedom.
If we look at the ground state of the quantum mechanical model with some fixed large $N$, the system already has
a particular Fermi energy, and in particular the choice of $N$ and $\epsilon_F$ is such that the system is either
close to the double scaling limit or not.  Thus we need to modify the model so that the quantum
mechanical behavior, and thus the eigenvalue repulsion, is turned on as the energy spreads among eigenvalues.
The dynamics must be such that the quantum mechanical behavior is an emergent phenomenon.

To see how to implement this, we start with the following Lagrangian, where for the moment we consider classical 
degrees of freedom:
\begin{equation}
L = {\rm Tr} \left(\frac{1}{2} \dot{M}^2 + \frac{1}{2} M^2 +\frac{g}{3} M^3 \right)
\quad ,
\end{equation}
where $M$ is a hermitian $N \times N$ matrix.  We diagonalize this matrix by writing $M = \Omega^\dagger \Lambda \Omega$,
where $\Lambda$ is the diagonal matrix of eigenvalues $\lambda_i$ and $\Omega$ is a unitary matrix.
The Lagrangian now becomes, after introducing the hermitian matrix $\dot{A} = -i \dot{\Omega} \Omega^\dagger$,
\begin{equation}
L =  \sum_{i=1}^N \left( \frac{1}{2} \dot{\lambda}_i^2 + \frac{1}{2} \lambda_i^2 + \frac{g}{3} \lambda_i^3 \right) + \frac{1}{2} \sum_{i<j}^N | \dot{A}_{ij}|^2 (\lambda_i - \lambda_j)^2
\quad .
\end{equation}
Finally, it will be more convenient to pass to the Hamiltonian description. Introducing the momenta $p_i$ conjugate
to $\lambda_i$ and $\Pi_{ij}$ conjugate to $A_{ij}$, we find
\begin{equation}
H = \sum_{i=1}^N \left( \frac{1}{2} p_i^2 - \frac{1}{2} \lambda_i^2 - \frac{g}{3} \lambda_i^3 \right) + \sum_{i<j} ^N \frac{|\Pi_{ij}|^2}{(\lambda_i - \lambda_j)^2}
\quad .
\end{equation}
Clearly it is the last term that controls the eigenvalue repulsion.  Equally clear, however, is that with this 
Hamiltonian the $\Pi_{ij}$ momenta are constants of the motion.  Thus we will have to modify the classical dynamics
if we wish to have the eigenvalue repulsion turn on dynamically.

More concretely, let us imagine that in our initial configuration, all but one of the eigenvalues are collected
very close together in the metastable minimum at $\lambda_i = -1/g$, and that we have one eigenvalue at rest
at $\lambda_1 \ll - N^{1/3}$.  The initial eigenvalue displacement is chosen so that the initial energy is larger than $N/g^2$, which is the order of magnitude of the energy of the final, filled Fermi sea.  For the bulk of the eigenvalues to be able to stay
near the bottom, the momenta $\Pi_{ij}$ will have to begin very close to zero.  

Note that the force between
two eigenvalues falls off very rapidly, with the inverse third power of their separation.  To understand the eigenvalue
dynamics with fixed $\Pi_{ij}$ more clearly, let's look for a moment at two eigenvalues in the absence of
an external potential.  If we choose the initial conditions (at large negative $t$) so that the left eigenvalue is moving to the right with
initial velocity $v$, $ \lim_{t \rightarrow -\infty} \lambda_l = v t$, while the right eigenvalue is initially at rest at  zero, $ \lim_{t \rightarrow -\infty} \lambda_r= 0$, the equations of motion are easily solved to give
\begin{eqnarray}
\lambda_l = \frac{1}{2} v t - \sqrt{ \left(\frac{1}{2} v t \right)^2 + \left(\frac{\Pi}{v} \right)^2} \\
\lambda_r = \frac{1}{2} v t + \sqrt{ \left(\frac{1}{2} v t \right)^2 + \left(\frac{\Pi}{v} \right)^2} \quad ,
\end{eqnarray}
where $\Pi$ is the ``off-diagonal'' momentum controlling the repulsion, as above.  Thus we see that as $t$ becomes large and positive, 
the left eigenvalue is now at rest at $\lambda_l = 0$, while the right eigenvalue is moving to the right with
velocity $v$.  Up to exchange of the eigenvalues, it looks simply like the free propagation of two eigenvalues through each other,
apart from a short period of time when the free eigenvalue separation $|v t|$ is small compared to $\Pi/v$.  
Note for future reference that the distance of closest approach for the two eigenvalues is simply given by $2 \Pi/v$.

With the above Hamiltonian, the initially excited eigenvalue would fall toward the rest of the eigenvalues,
ultimately transferring the eigenvalue momentum to the rightmost eigenvalue, which would
subsequently escape into the instable region at positive $\lambda$.  All of this would happen on a time scale
of order $1/\sqrt{N}$.
We see that in addition to needing the $\Pi_{ij}$ to vary, they need to be able to absorb some of the energy,
and thus thermalize the system, on a similar time scale.
Further, we require that this new thermalization mechanism only acts on ``distance'' scales $\lambda_i-\lambda_j < \delta$, for some fixed $\delta$ of order $1/Ng$.
This is required both so that we can initially separate one eigenvalue that contains all the energy, and also
so that we can end up with something like free propagation for eigenvalue perturbations after equilibrium is reached 
and the eigenvalues fill out the fermi sea.  To see this, note that once the $\Pi_{ij}$ have grown somewhat, and very high
momentum eigenvalues have escaped to the region of positive $\lambda$, that the distances of closest approach,
of order $\Pi_{ij}/v$ will have some lower bound, which will ultimately be greater than or equal to $\delta$. 

We would like to comment at this point that 't Hooft has argued \cite{'tHooft:1999gk} that reconciling holography
with ideas of locality suggest a picture similar to what we have here. Namely, we find a system with classical, deterministic
degrees of freedom, which has a short distance effect that both loses information (here through thermalization) and 
ultimately produces quantum mechanical effects (here corresponding to the growth and stabilization of the
eigenvalue repulsion).  We shall discuss this more in the final section. 

\section{From Eigenvalues to Spacetime}
We begin this section by summarizing the picture so far.  We begin with all the $\Pi_{ij}$ very close to
zero, so that all but one of the eigenvalues can sit at the bottom of the metastable minimum, at $\lambda_i = - 1/g$.
In particular, we require that the initial spread between eigenvalues is much smaller than the special distance
$\delta$ introduced above, although the total spread, since $N$ is assumed very large, may be larger than $\delta$.  The final eigenvalue begins at rest at an initial negative displacement, large
on the scale of $N^{1/3}$, so that it's energy is large on the scale of $N/g^2$.  The eigenvalue falls down, and
as it approaches within the distance $\delta$ of the leftmost eigenvalues, a subset of the eigenvalues starts
to thermalize, spreading out the energy by increasing some of the $\Pi_{ij}$ momenta.  Although the dynamics of
our model ensures that the eigenvalues remain in the same order throughout, up to eigenvalue permutation we can follow
the trajectory of our special excited eigenvalue.  As it travels through the sea of other eigenvalues, it keeps
shedding energy through thermalization.  Since we imagine that the thermalization occurs while eigenvalues are within
a distance $\delta$, the process of thermalization will have some non-trivial behavior.  In particular, initially
the excited eigenvalue will have a higher velocity, and so will spend a smaller amount of time very close to 
the leftmost eigenvalues, while it will spend more time very close to the rightmost eigenvalues.  Therefore, in
the initial stages of excitation, there will be {\em more} energy deposited on the right-hand part of the spectrum.
Whether this tilt persists all the way into equilibrium will clearly depend on how quickly it is reached.

Although so far our system is still very far from equilibrium, to trace the behavior further, it is helpful
to look at the equilibrium dynamics of the system, where we now allow the $\Pi_{ij}$ momenta to freely vary.
The partition function for this system is
\begin{eqnarray}
Z &= &{\rm Tr} e^{- \beta H} \nonumber \\
&= &\int \prod_{i=1}^N (d\lambda_i dp_i) \prod_{i<j}^N (dA_{ij}d\Pi_{ij} )
\exp{- \beta  \left[\sum_{i=1}^N \left( \frac{1}{2} p_i^2 - \frac{1}{2} \lambda_i^2 - \frac{g}{3} \lambda_i^3 \right) + \sum_{i<j} ^N \frac{|\Pi_{ij}|^2}{(\lambda_i - \lambda_j)^2} \right]}
.
\end{eqnarray}
This partition function is straightforward to evaluate.  The integral over $p_i$ contributes an $N$ dependent
constant times $\beta^{-N/2}$, while the integral over $A_{ij}$ simply contributes an $N$ dependent volume
factor.  The integral over $\Pi_{ij}$ provides a factor of $\beta^{-N(N-1)/2}$, and also has the effect of introducing the square of the Vandermonde determinant,
$\Delta(\lambda) = \prod_{i<j}^N (\lambda_i - \lambda_j)$, into the measure for the remaining $\lambda_i$.
Our partition function has thus become
\begin{equation}
Z = C_N \beta^{-\frac{1}{2}N^2} \int \prod_{i<j}^N d\lambda_i \Delta^2(\lambda) \exp{ \beta \sum_{i=1}^N \left( \frac{1}{2} \lambda_i^2 + \frac{g}{3} \lambda_i^3 \right)}
\quad ,
\end{equation}
with $C_N$ an $N$ dependent constant.  The integral in the above expression is simply the path integral for the
zero-dimensional (i.e. no time coordinate) quantum matrix model.  Now we find ourselves in very familiar territory.
Following the classic treatment of Brezin, Itzykson, Parisi, and Zuber \cite{Brezin:1977sv}, we can estimate this
integral, in the large $N$ limit, by doing a saddle point approximation.  The result, after translating to
our conventions, gives both an expression for the free energy (whose specific form will not concern us
here) and an equilibrium distribution for the eigenvalues given by
\begin{equation}
\rho(\lambda) = \frac{\beta}{2 \pi N} (\sigma - g \lambda) \sqrt{(\lambda - \lambda_i)(\lambda_+ - \lambda)}
\quad ,
\end{equation}
where the turning points are located at
\begin{equation}
\lambda_{\pm} = - \frac{\sigma+1}{g} \pm  2 \sqrt{\frac{N}{\beta(1 + 2\sigma)}} \quad ,
\end{equation}
and $\sigma$ is a solution of the equation
\begin{equation}
\frac{2 g^2 N}{\beta} + \sigma (1 + \sigma) (1 + 2 \sigma) = 0 \quad .
\end{equation}
Note that for our choice of potential, the local maximum is located at $\lambda = 0$,
with the metastable minimum at negative $\lambda$ and the instable region at positive
$\lambda$.
For our model, the eigenvalues will also have a gaussian distribution of momenta superposed over the above distribution.
If we ignore for the moment the complications of non-equilibrium dynamics, we note
the following feature.  As more energy is deposited into the eigenvalue sea, it's temperature
rises, and thus $\beta$ will be decreasing.  There is a critical point in the behavior at 
$\beta = 12 \sqrt{3} g^2 N$. At this point $\lambda_+ = \sigma/g$ and in particular
the right edge of the distribution will now go as $(\lambda_+ - \lambda)^{3/2}$.
We expect this critical point to be related to the double scaling limit in our model.
In particular, the critical point in the usual matrix model is precisely the point at 
which you expect to get string dynamics, and is also the point when you expect eigenvalues
to start spilling over towards the instability.

Returning the discussion to the non-equilibrium dynamics, we will assume that the
energy transferred to the sea by the excited eigenvalue is greater than that needed
to reach the critical point.  The effect of the dynamics will now be to have a tilt,
as discussed above, which will affect the momentum distribution of the eigenvalues.
Further, we expect that any eigenvalues with large enough momenta will escape to
positive $\lambda$, so that as equilibrium is approached we have some smaller number of
eigenvalues present.  

As long as the friction induced by the short distance thermalization
is great enough, the sea will end by filling up the metastable dip, with excess
eigenvalues draining off.  Here we notice another interesting feature of our toy model.
In the usual quantum mechanical theory, the instability of the bosonic string is
a problem, and more precise statements require moving to the Type 0B string, which has a
quartic potential and is thus stable.  For our scenario, the world-sheet supersymmetric
string is of no help, as having a quartic potential would require an exquisite
fine tuning of the initial conditions.  The instability of the bosonic string is
the precise feature that allows our system to dynamically approach the continuum limit.

Let us also note the following: the higher the effective friction acting on the
excited eigenvalue, the more uniformly we might expect the eigenvalue ``soup'' to
heat up.  If it heats up more uniformly, there should be smaller fluctuations in the momenta
and thus less ``sloshing'' of eigenvalues over the edge.  This will in turn result
in a smaller $\mu$ for the final state.  Noting further that $\mu$ is proportional
to one over the string coupling ($\mu \propto 1/g_s$),
we find that strong friction is correlated with strong string coupling.
 
Now we would like to comment on the cosmology seen by a string observer, after
the system is already sufficiently close to the double scaling limit.  In the final
stages of thermalization, the sea will still be expanding somewhat.  Thus an observer
will see some degree of spacetime expansion.  Since this expansion is fueled by the
thermalization, and not by normal ``quantum" dynamics, an observer may postulate
that during this period there is an exotic source of stress-energy, which however
will be absent at late times.  In addition, the tachyon, which is given by fluctuations
of the fermi surface, will have some characteristic spectrum of fluctuations.
The details of this spectrum will depend on the initial energy, the strength of
the friction caused by thermalization, the behavior of the phase transition, and
the subsequent draining of highly excited tachyons.  However, we may note here 
that the phase transition may produce a scale invariant spectrum, while the
non-equilibrium dynamics may superpose a tilt, as sketched previously.  We may further expect that very violent
fluctuations will result in eigenvalues that separate from the collective sea.  From the point of view of the
string theory background, these highly excited eigenvalues should correspond to unstable D-branes.
In principle,
this system is simple enough that this may be studied numerically.  However, it
is not clear at this point if having a more precise answer would help our understanding
of the scenario as applied in higher dimensions.

\section{Discussion: Lessons for Higher Dimensions}
We will conclude by summarizing what we've learned, and what import it may have for a more realistic, higher-dimensional
model.  Most importantly, we have seen that it is possible to implement our scenario in this simplified context;
an engine can be set up, operating between a small number (here one) of highly excited degrees of freedom and a
large number of frozen degrees of freedom, which drives the system to expand into a stringy phase.  We have also
been able to extract some insight into the dynamics that should determine the tachyon fluctuation spectrum; this
 should help in understanding the production of density perturbations in a realistic scenario.  In particular, we
 see that there are two important features which are natural to understand first.  The first is what the order of
 the phase transition is expected to be, and in particular whether such a transition produces a scale invariant
 spectrum.  The second feature to understand is in {\em which} direction the non-equilibrium dynamics will tilt
 the spectra.  If these two properties can be robustly understood, then we may be able to falsify the picture through
 comparison with the cosmic microwave background.

One feature of our toy model that seems rather surprising is that we have had to alter the model so that quantum
mechanics is an effective phenomenon that emerges at late times.  It is possible that this modification is made
necessary by the special kinematics of 2D.  In particular, since the extent of the fermi sea is determined simply
by fermionic eigenvalue repulsion, the only way for the ``universe" to expand is to have the eigenvalue repulsion,
and thus $\hbar$, growing in time.  It is, however, entirely possible that such a modification must hold for a 
higher dimensional scenario.  't Hooft has argued  \cite{'tHooft:1999gk} that reconciling conventional
ideas of locality with the apparently non-local behavior for quantum gravity implied by holography may be resolved
in a similar manner.  In particular, he imagines a classical theory for gravity that is conventionally local.  This
classical theory, however, has some dissipative dynamics on short scales which lead to information loss. Quantum
mechanics, and the reduced entropy implied by holography, arise through the statistical mechanical treatment of equivalence
classes of states, where two states are in the same class if they both evolve to a common state in the future.
Our toy model displays a similar behavior: information is lost through the short distance thermalization that allows
the off-diagonal momenta $\Pi_{ij}$ to vary.  We find this prospect very exciting.  If the higher dimensional model
{\em does} turn out to require such modification, then it may be fruitful to return once again to our toy model.
In this case, important insight may be gained by understanding the specifics of possible thermalization processes, as well
as more precisely understanding the emergence of quantum mechanics.

Finally, we point out one more interesting feature of our model.  Our scenario required the bosonic string, with
it's associated cubic instability, in order to dynamically approach the double scaling limit.  This could indicate that supersymmetry,
while important for controlling quantum fluctuations in the critical string, has no preferred role in our scenario.  
Indeed it may be the case that apparent instabilities provide critical elements to the dynamics.  In this context,
it may be interesting to consider more general non-critical and non-supersymmetric string theories \cite{future}.

\end{document}